\documentclass[twoside,a4paper]{article}
\usepackage{amsmath,graphicx,amssymb,fancyhdr,amsthm,,textcomp,verbatim} 
\usepackage{authblk}
\usepackage{soul} 
 
\usepackage[usenames]{xcolor}

\theoremstyle{definition} 
  
\theoremstyle{remark}  
  
\def\beq{\begin{eqnarray}}  
\def\eeq{\end{eqnarray}}  
\def\bsp{\begin{split}}  
\def\esp{\end{split}}

\begin{document}  

\title{Geometric Horizons in Binary Black Hole Mergers} 

\author[1]{\textsc{Alan Coley}}
\author[1,2,3]{\textsc{Jeremy M. Peters}}
\author[2,3,4]{\textsc{Erik Schnetter}}
\affil[1]{Department of Mathematics and Statistics, Dalhousie University, Halifax, Nova Scotia, B3H 3J5, Canada}
\affil[2]
{Perimeter Institute for Theoretical Physics, Waterloo, Ontario, N2L 2Y5, Canada}
\affil[3]{Department of Physics \& Astronomy, University of Waterloo, Waterloo, ON N2L 3G1, Canada}
\affil[4]{Center for Computation \& Technology, Louisiana State University, Baton Rouge, LA 70803, USA}
  
\date{\today}  
\maketitle  
\pagestyle{fancy}  
\fancyhead{} 
\fancyhead[EC]{}  
\fancyhead[EL,OR]{\thepage}  
\fancyhead[OC]{}  
\fancyfoot{} 

\begin{abstract}

We numerically study the algebraic properties of the Weyl tensor 
through the merger of two non-spinning black holes. We are particularly interested
in the  conjecture that for such a vacuum spacetime, which is zeroth-order algebraically general, a geometric horizon, on which the spacetime is algebraically special and
which is identified by the vanishing of a complex scalar invariant (${\mathcal{D}}$),
characterizes a smooth foliation independent surface (horizon) associated with the black hole.
In the first simulation we investigate the level--$0$ sets of $\text{Re}({\mathcal{D}})$
(since $\text{Im}({\mathcal{D}})= 0$) in the
head-on collision of two unequal mass black holes. In the second  simulation we shall investigate the level--$\varepsilon$ sets of $|{\mathcal{D}}|$
through a quasi-circular merger of two non-spinning, equal mass black holes. The numerical
results, as displayed in the figures presented, provide evidence that  
a (unique) smooth geometric horizon can be identified throughout all stages of the  binary black hole merger.

\end{abstract}

\newpage

\section{Introduction}

The event horizon of a black hole (BH) solution in General Relativity
(GR) is defined as the boundary of the non-empty complement of the
causal past of future null infinity; i.e., the region for which signals sent from the
interior will never escape \cite{AshtekarKrishnan}.
To examine the interaction of realistic BHs with their environment in
numerical GR \cite{T}, in the 3+1 approach and/or in the Cauchy-problem in GR, it is
necessary to locate a BH quasi-locally \cite{AnderssonMarsSimon,Jaramillo}. A local characterization may not
rely on the existence of an event horizon alone, as BHs are expected to
undergo evolutionary processes and are typically dynamical.

The concept of closed trapped surfaces without border, which are compact spacelike surfaces such that the expansions of the
future-pointing null normal vectors are negative, was proposed in \cite{Penrose:1964wq}.  A related concept to trapping surfaces are marginally outer trapped surfaces
(MOTSs) which are two-dimensional (2D) surfaces for which
the expansion of the outgoing null vector normal to the
surfaces vanishes. In more detail, let $\Sigma$ be a compact spacelike 2D surface without border, and consider light rays leaving and entering $\Sigma$, with null directions $l$ and $n$, respectively.  Let $q_{ab}$ be the induced metric on $\Sigma$ and denote the respective expansions as $\Theta_{(l)} = q^{ab}\nabla_al_b$ and $\Theta_{(n)} = q^{ab}\nabla_an_b$ \cite{news}.  Then, $\Theta_{(l)}$ and $\Theta_{(n)}$ are quantities which are positive/negative if the null rays locally diverge/converge.  
$\Sigma$ is called a {\it closed trapped surface} if $\Theta_{(l)}<0$ and $\Theta_{(n)}<0$ \cite{Penrose:1964wq,ERIK}.  $\Sigma$ is a (future) marginally outer trapped surface (MOTS) if it has zero expansion for the outgoing light rays, $\Theta_{(l)} = 0$ (and $\Theta_{(n)} < 0$) \cite{ERIK,Schnetter2020_2,Schnetter:2006yt}.

The outermost MOTS is called the apparent horizon (AH). 
Assuming a smooth time evolution for the MOTSs,
the 2D surfaces can be combined to construct a 3D surface
known as a marginally trapped tube (MTTs) \cite{BoothFairhurst}. If the MTT is foliated by MOTSs then it is called a dynamical horizon 
(DH) \cite{AshtekarKrishnan,BoothFairhurst}.
Unlike the event horizon, AHs are quasi-local
and they  are highly
non-unique  \cite{AG2005}. Although AHs are unique within a spacelike hypersurface,
without a particular choice of hypersurface there are ”too many AHs” to be
useful. The mathematically elegant construction ”union of all trapped surfaces”
or ”outermost marginally trapped surface” are not constructive and are thus
not useful in practice. MTTs and DHs are 
intrinsically foliation-dependent, and hence observer dependent.
It is crucial to locate a BH locally. A DH is particularly well-suited to analyze realistic dynamical processes involving BHs
such as BH growth and coalescence \cite{AshtekarKrishnan,Senov}. 
The above definitions serve as a quasi-local description of BHs \cite{CMS,Evans}.  
One disadvantage of AHs is that the AHs observed depend on how the spacetime is foliated, so that AHs are observer dependent \cite{AG2005}.

\subsection*{The Geometric Horizon  Conjecture}

A foliation invariant quasi-local and more geometrical approach, which is an alternative approach to MOTS, has been proposed to discuss BH spacetimes \cite{CMS}.
Since we are primarily interested in numerical applications in 4D
to study the (asymmetric) collapse or merger of real BHs, the relevent spacetimes are of general algebraic type away from the horizon.
The {\em{geometric horizon conjecture}} states that \cite{CMS}:
{\em{If a BH spacetime is zeroth-order algebraically general, then on the geometric horizon (GH) the spacetime
is algebraically special.}} 
Hence a GH can be identified by the alignment
type {\bf II} or {\bf D} ({\bf II/D}) discriminant conditions in terms of
scalar (curvature) polynomial invariants (SPIs) \cite{class}; that is, a particular
set of SPIs  vanish on the GH, thereby defining the {\em{geometric horizon}}  \cite{CMS}.

We note that $SPI$s may not specify the GH completely in the sense that they may also vanish at fixed points
of any isometries and along any axes of symmetry. Unlike 
AHs, we (again) remark that a GH does not depend on a chosen foliation in the
spacetime \cite{AG2005}.  
In physical problems with dynamical evolution the 
horizon might not be unique, or may not exist at all, and amendments to the conjecture may be necessary
(e.g., it may be appropriate to replace the vanishing of invariants in the definition of a 
geometric horizon as an algebraically special hypersurface, with the conditions that the                            
magnitudes of certain $SPI$s take their smallest values).

In this Letter we wish to evaluate this conjecture by studying the GH  numerically
in two physically relevant  situations: (i) the head-on collision of unequal mass BHs, (ii) 
two merging, equal mass and non-spinning BHs, during a 4D binary BH (BBH) coalescence  \cite{kramer}.
In particular, we shall locally determine the foliation invariant GH based on the algebraic (or Petrov) classification of the Weyl tensor \cite{CMS}. 
On the GH, the Weyl tensor is algebraically special.
We can use discriminants to determine the necessary conditions for this in terms of simple
$SPI$s,  utilizing  the algebraic classification (using the boost weight decomposition \cite{class}) of the Weyl and Ricci tensor when treated as curvature operators \cite{BIVECTOR}, for the spacetime  to be of special algebraic type {\bf II} or {\bf D}. 
The necessary real conditions for the Weyl tensor to be of type {\bf  II}/{\bf D} are given in \cite{CH}.
These 2 real conditions are equivalent to the vanishing of the real and imaginary parts of the complex invariant ${\mathcal{D}} \equiv I^3-27J^2=0$ in terms of
the complex Weyl spin tensor in the Newman-Penrose (NP) formalism \cite{kramer}.
Alternatively we can use the discriminant analysis to provide the type {\bf II/D} syzygies expressed in terms of 
$SPI$s by treating the Weyl tensor as a trace-free operator acting on the 6-dimensional vector space of bivectors \cite{CH}. 
More practical {\it necessary} conditions can be obtained by
considering the eigenvalue structure of the trace-free symmetric part of the operator
$C_{abcd}C^{ebcd}$. This can also be applied to study where
the covariant derivative of the Weyl tensor ($ \nabla W$) 
is of algebraically special type {\bf II/D}.

The two applications are complementary. In the head-on collision, the complex part of   ${\mathcal{D}}$ vanishes, so that ${\mathcal{D}} \equiv {\mathcal{D}_r}$, where ${\mathcal{D}}_r = \text{Re}({\mathcal{D}})$, is not of a single sign and hence can be used to locate zeros.
Unfortunately the numerics in this case is not always sufficient for conclusive results.
In the second application ${\mathcal{D}}_i \neq 0$.
The complex invariant ${\mathcal{D}}$ vanishes if and only if its magnitude, $|{\mathcal{D}}| = 0$.  However, numerically finding the level--$0$ sets of $|{\mathcal{D}}|$ precisely is extremely difficult because $|{\mathcal{D}}|$ is positive definite.
In practice, due to numerical resolution and numerical errors,
instead of analyzing the level--$0$ sets of ${\mathcal{D}}$, we shall consider the level--$\varepsilon$ sets for small $\varepsilon$ (e.g., $\varepsilon\in\left\{3\times 10^{-4},\;5\times 10^{-4},\;1\times 10^{-3}\right\})$.

We wish to study the behaviour of the complex ${\mathcal{D}}$
(and particularly $|{\mathcal{D}}|$), through a BBH merger.  Since the Kerr geometry is type ${\bf D}$ everywhere, it follows that ${\mathcal{D}} = 0$ everywhere for a Kerr BH.  It is known that in a BBH merger the merged BHs at late times settle down to a solution well described by the Kerr metric \cite{CMS}.  Thus, for a merger of 2 initially Kerr BHs, a plot of the real part and imaginary part of ${\mathcal{D}}$ should be roughly zero everywhere at early and late times (and simply give rise to random (numerical) noise);
in particular, the numerical noise (artificial zeros) make early time plots somewhat unreliable. However, in the intermediate ``dynamical" region (during the actual merger and coalescence at intermediate times) the Petrov type
is ${\bf{I}}$, and the zeros of ${\mathcal{D}}$ will give us important information.

The horizon of the exact Kerr solution is, in fact, located by the condition that 
$\nabla W$ is of type {\bf{II/D}} there (characterized by the fact that 
various SPIs of $\nabla W$ are zero) \cite{GANG}. 
The second part of the GH conjectures states that if a BH spacetime is algebraically special (so that on the GH the BH spacetime is automatically algebraically special) and if, for example, $ \nabla W$ is algebraically general, then on the GH, $\nabla W$ is algebraically special \cite{CMS}.  
Therefore, it might be useful to also study the
algebraic type of $\nabla W$, particularly at early and late times. This might be best done via the NP formalism and the use of appropriate Cartan invariants.

Indeed, in addition to algebraic and differential $SPI$s, Cartan scalar invariants can also be defined and used \cite{GANG}.  More specifically, for a fixed set of frame vectors a {\it Cartan invariant} is a scalar that is constructed from the Weyl or Riemann tensor or any of its covariant derivatives by contracting the Weyl tensor (resp., any of its covariant derivatives) with the frame vectors.  Cartan invariants are easier to compute in general than $SPI$s (because, for example, they are linear in the Weyl tensor and its covariant derivatives).  
It has been found that in a ``preferred frame" there is  evidence that the NP spin coefficients $\rho, \mu$, which are related to
$\Theta_{(l)}$, and $\Theta_{(n)}$,  can be identified with the surfaces on which $\nabla W$ is algebraically special; that is, 
the GH surfaces are characterized by the conditions that the spin coefficients $\rho, \mu$ are zero \cite{CMS,CMKT,szek}.
Indeed, the spin coefficients $\rho$ and $\mu$
were utilized in the numerical studies of 
\cite{CMS} and \cite{CMKT}. 
In future we could further investigate $\nabla W$
or the spin coefficients $\rho, \mu$ (in the prefered frame)
in the current applications,
but this will present some numerical and analytic difficulties.
In addition, the question of whether these spin coefficients
take on a special meaning in a ``preferred frame'' will be further investigated elsewhere.

\subsubsection*{Examples and motivation}

There are many examples  \cite{GANG} that 
support the GH conjectures. 
There is also motivation for the conjectures
from analytical results
\cite{AshtekarKrishnan}.
For example, in 4D, and assuming that the dominant energy condition holds, it was shown that on a non-expanding weakly isolated null horizon
the Ricci and Weyl tensors are of type {\bf II/D}  \cite{Ashtekar},
and that $\nabla W$ is of  type {\bf II} (on the horizon) \cite{CMS}.

The conjecture is intended to apply to DHs, which are     
more difficult to study.
There are examples of dynamical BH solutions that admit GH
\cite{CMS}, such as dynamical BH solutions which are conformally
related to stationary BH solutions and the imploding spherically
symmetric metrics [17]. 
For these particular dynamical BH examples, the GHs correspond to MTTs.
However, in general a GH will not be a MTT, as the preferred null direction
will not necessarily be geodesic and surface forming. 
In addition, for a
given mass function, the Vaidya spacetime also provides explicit examples of the
transition from the dynamical to isolated horizons  \cite{kramer}.

It has been demonstrated that spherical symmetric dynamical BH solutions
must admit a unique, invariantly defined dynamical GH \cite{CMS}. Non-spherically symmetric BH solutions
have also been considered. Quasi-spherical 
{Szekeres} dust models, which can act as physical models for the formation of  primordial BHs when the BH's mass and
collapse time are fine-tuned in order to avoid shell-crossings forming outside of the AH,
are known to admit an AH and it was
shown \cite{szek} that this hypersurface is, in fact, a GH \cite{CMS}.
In addition, Kastor and Traschen (KT) have found a family of exact closed universe solutions to the Einstein-Maxwell equations with a cosmological constant representing an arbitrary number
of charged Q = M BHs \cite{KT}.
In the dynamical two-black-hole KT solution \cite{KT}, 
when the sum of the two BH masses does not exceed a critical mass, the
BHs coalesce and form a larger BH. Subsequently,
the existence of invariantly defined quasi-local GH hypersurfaces 
in the KT solution containing multiple BHs 
was demonstrated  \cite{CMKT}. 
In particular, the existence of GHs in the
three equal mass BH KT solution was also  shown in \cite{CMKT}.
Some aspects of the multiple BH
KT solutions were investigated in \cite{KTnsh}.

\subsubsection*{Numerics}

In numerical relativity it is most useful to utilize an initial value formulation of GR (using a $1+3$ approach), where initial data is specified on a Cauchy hypersurface and is then evolved forward in time, which necessitates a local description of BH solutions \cite{T,AnderssonMarsSimon,Jaramillo}.  Indeed, in numerical studies of time-dependent collapse, it is often more practical to track AHs \cite{Booth2005}.
Gravitational fields at the AH are correlated with gravitational wave signals \cite{Evans,6}, so AHs are useful to study (simulations of high
precision waveforms of) gravitational waves.  AHs are also used in numerical simulations of BBH mergers and the collapse of a star to form a BH \cite{Booth2005}.   
The recent observations by the LIGO collaboration of gravitational waves
from BBH mergers make use of the comparison with templates
in which AHs  play an important role \cite{LIGO}.
In addition, MOTSs turn out to be well-behaved numerically, and can be used to trace physical properties of BHs as they evolve over time and through a BBH merger \cite{ERIK,Schnetter:2006yt}.

The Einstein toolkit \cite{Loffler:2011ay}
infrastructure was used for the numerical calculations used in this paper. In \cite{ERIK} a new  horizon finder utilizing the  stability operator for MOTS \cite{AnderssonMarsSimon,Andersson_2009}
for locating MOTSs numerically was presented.
The novelty of  \cite{ERIK} is that it uses a much more advanced discretization of the
horizon surface and it can find not only the outer most MOTS (i.e., the AH)
but also the 3 other MOTS that are nested inside the AH at late times in BBH
mergers; this horizon finder is therefore capable of finding even very highly distorted MOTSs and it is possible to track the computation up to the merger
point, and beyond, and hence better understand the merger process.

In the simulation of the head-on collision of two unequal mass  non-spinning BHs
axisymmetry is enforced
and $6$th order finite differencing on a uniform grid is used, and
Brill-Lindquist 
initial data is considered, representing a BBH system at a
moment of time-symmetry. 
We note that numerical resolution is lost at
later times when the horizons get too close to the punctures. 
In the simulation of two merging, equal mass and non-spinning BHs, the simulations are run using $4^{\text{th}}$ order finite differencing on an adaptive mesh grid, with adaptive refinement level of 6 \cite{AMR}. In addition, Brill-Lindquist initial data with BH positions and momenta set up to satisfy the ``QC-0'' initial condition \cite{QC0} is utilized.

\newpage

\section{Head-on collision of unequal mass BHs}

The merger of two (binary) BHs (BBHs) is often visualized by an event
horizon; an example of this is
\cite{Matzner:1995ib}, showing the well known ``pair of pants''
picture for a BBH collision \cite{ERIK,Matzner:1995ib}.  
However, event horizons are
not generally suitable for extracting quantities of physical interest
and tracking them all the way through the merger in quantitative studies \cite{Penrose:1964wq}. 
It is known that MOTSs and AHs can be used to determine and track the time evolution of physical properties of a BH, such as such as mass and angular momentum \cite{news,ERIK,Schnetter:2006yt,Evans,6,12}.

In both \cite{ERIK} and \cite{Evans}, the head-on collision of two non-spinning unequal mass BHs was numerically simulated, and strong numerical evidence for 
the merger of MOTSs was found.  
In \cite{ERIK}, it was found that there is a connected sequence of MOTSs, which interpolate between the initial and final states of the non-linear merger (two separate BHs to one BH, respectively) \cite{ERIK}.  The MOTS were found using a horizon finder \cite{AnderssonMarsSimon,ERIK,Andersson_2009}.  This interpolating sequence of MOTSs allows physical BH quantities to be traced through the merger.  In \cite{Evans,PhysRevLett.78.3606}, the dynamics of the head-on collision was studied by modelling the two initial BHs as spacetime punctures and the initial BH separations and mass ratios were varied.  The AHs of the initial BHs in the simulations were used to track the location and properties of the BH punctures \cite{Evans}.  

Both \cite{ERIK} and \cite{Evans} have described the structure of the MOTSs through the merger in detail.  Initially, there are two BHs with disjoint MOTS (which are AHs at this point \cite{Evans}), 
$\mathcal{S}_1$, and $\mathcal{S}_2$, one around each BH
(and $\bar{\mathcal{S}}_1$, and $\bar{\mathcal{S}}_2$ to describe their spherical approximations).  
It was noted that if the initial separations are small enough and the lapse function is properly behaved, then the two initial MOTS $\mathcal{S}_1$ and $\mathcal{S}_2$ are approximately null surfaces \cite{Schnetter:2006yt}, and hence isolated horizons \cite{Evans}.
Then, at the time of the merger, a common MOTS forms around the two separate BHs and bifurcates into an inner MOTS, $\mathcal{S}_i$, which surrounds the MOTS and an outer MOTS, $\mathcal{S}_c$.  The outer MOTS $\mathcal{S}_c$, increases in area, encloses the three inner MOTS, $\mathcal{S}_1$, 
$\mathcal{S}_2$, and $\mathcal{S}_i$, is the AH of the common BH after the merger has taken place \cite{ERIK,Evans}.  The fate of this common AH is well understood \cite{Schnetter:2006yt,Evans,4}.  The inner MOTS, $\mathcal{S}_i$, decreases in area and approaches the inner MOTS $\mathcal{S}_1$ and $\mathcal{S}_2$ \cite{ERIK,Evans}.  This bifurcation and the three inner MOTS, $\mathcal{S}_1$, $\mathcal{S}_2$, and $\mathcal{S}_i$ have also been well studied \cite{ERIK,Schnetter:2006yt,Evans,12,4,13}.  In \cite{Evans}, it was found that at late times, $\mathcal{S}_1$ and $\mathcal{S}_2$ continued to exist and intersected in general but remained separate horizons.  The BH punctures, while acting effectively as a single puncture, did not completely merge \cite{Evans}.  In \cite{ERIK}, it was found that at the time when the inner surface touched the two individual BHs, the inner MOTS displayed self-intersections and thus a topology change, and a slight area increase.


\begin{figure}
    \centering
    \includegraphics[width = 11cm]{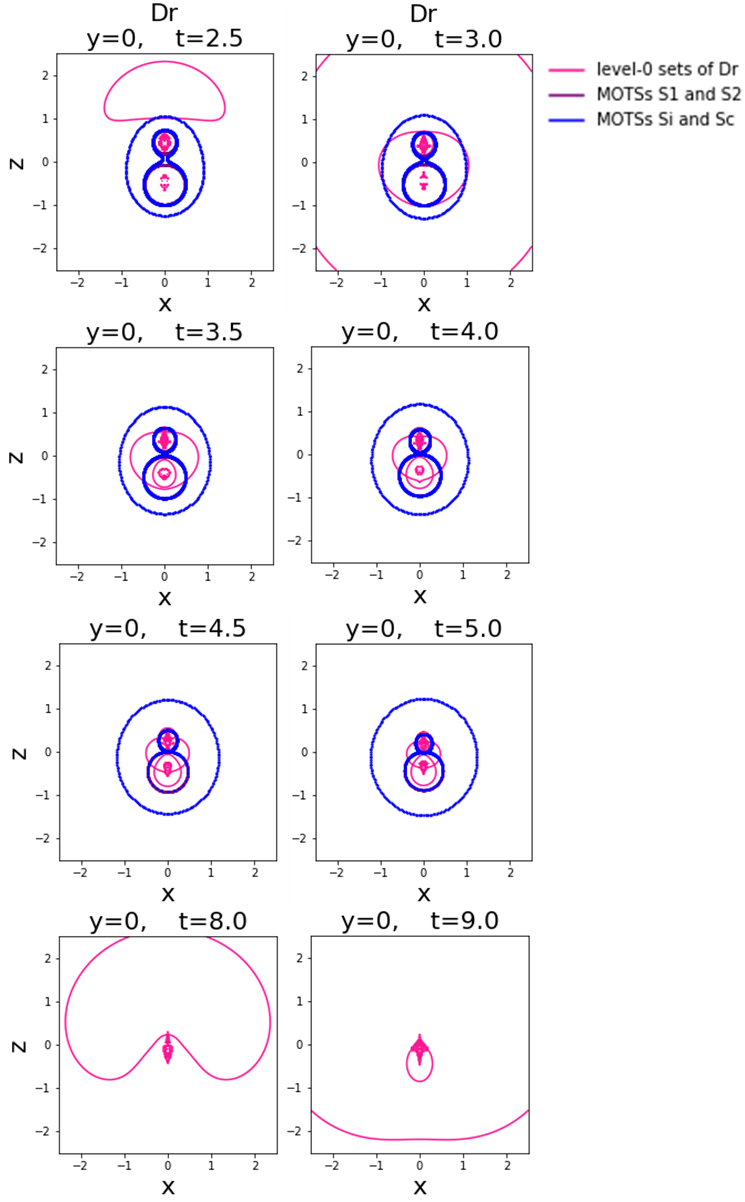}
    \caption{Plots of the level--0 sets of ${\mathcal{D}}_r = \text{Re}({\mathcal{D}})$ away from the $z$ axis at times $t = 2,\;3,\;3.5,\;4,\;4.5,\;5,\;8,\;9$, respectively, from the head-on axisymmetric collision of two unequal mass BHs.  The data for $\mathcal{S}_{1,2,i,c}$ were presented previously \cite{ERIK}; for more details on the interplay between $\mathcal{S}_{1,2,i,c}$ see figures 1-3 therein.}
    \label{HeadOn}
\end{figure}

\subsection*{Discussion}

We utilize the numerical data 
from \cite{ERIK}, based on the
Einstein toolkit and enforcing
axisymmetry and  starting with Brill-Lindquist 
initial data representing a BBH system at a
moment of time-symmetry,
for the head-on axisymmetric collision of two non-spinning unequal mass BHs with mass ratio
$\frac{m_{\{z>0\}}}{m_{\{z<0\}}} = \frac{0.8}{0.5}$.
In figure 1 we present the time plots for $t=0.5-5.5$
(and at later times t = 8, 9); by $t=1/2$ the outer MOTS has already formed. 
We lose numerical resolution at
later times when the horizons get too close to the punctures. 
Therefore, we did not look for MOTS there.

Since in this simulation
${\mathcal{D}}_i = \text{Im}({\mathcal{D}}) = 0$
(i.e.,  ${\mathcal{D}}_r \equiv \text{Re}({\mathcal{D}}) = {\mathcal{D}}$),
to study the level--$0$ sets of ${\mathcal{D}}$ it suffices to study the level--$0$ sets of ${\mathcal{D}}_r$. In particular, the
changing of sign through a zero of $D = D_r$ reveals a MOTS
(and there are no problems regarding resolution since we are not considering a positive definite quantity as in second application). 
In figure \ref{HeadOn}, we compare the interpolating sequence of MOTSs with the level--$0$ sets of ${\mathcal{D}}_r$. We plot ${\mathcal{D}}_r$ vs $x$ and $z$ for $y = 0$. We obtain the data for $y \neq 0$ by using cylindrical coordinates $(x,z,\theta)$ to describe the system, with $\theta$ describing rotations around the $z$ axis in the right-hand sense, so that the data for $\theta = \theta_0$ is obtained by rotating the data for $\theta = 0$ through the angle $\theta_0$.  

It should be noted that ${\mathcal{D}}_r = 0$ at many points on the $z$ axis, but these points have been omitted in figure \ref{HeadOn}.  Indeed, in \cite{CMS} it was noted that ${\mathcal{D}}$ could vanish additionally on axes of symmetry, which seems to be the case in this present simulation, as $z = 0$ is an axis of symmetry for the simulation, as described in the previous paragraph.  We also notice that ${\mathcal{D}}_r = {\mathcal{D}} = 0$ in a dense region in the interior of $\mathcal{S}_1$ and $\mathcal{S}_2$.  
As noted above, at early times the configuration is close to 2 individual Kerr BHs so that the spacetime is of type {\bf{D}} almost everywhere and so numerical noise (artificial zeros) make early time plots somewhat unreliable with no clear closed curves visible.

Aside from this region, the level--$0$ set of ${\mathcal{D}}_r$ is seen to be partitioned into $2$ simple closed curves at time $t = 3$.  
One such curve interpolates between
$\mathcal{S}_i$ and $\mathcal{S}_c$, while the other curve encircles $S_c$ at time $t = 3$.  
At $t = 3.5$, a third simple closed curve becomes visible, which lies in the interior of $\mathcal{S}_2$ (purple), lying in the component of $\mathcal{S}_i$  where $z\leq 0$.  Note that the purple curves 
$\mathcal{S}_1$ and $\mathcal{S}_2$ happen to be overlaid by the blue $\mathcal{S}_i$.  
Then at times $t = 4,\;4.5,\;5$, the two visible closed curves shrink so that they reasonably approximate $\mathcal{S}_i$ and $\mathcal{S}_2$ (overlaid by the bottom component of   
$\mathcal{S}_i$). (The top component of $\mathcal{S}_i$, which coincides with $\mathcal{S}_1$
(above $z=0$), is tracked by a dense region of the ${\mathcal{D}}_r$ level curve). 
However, one of these two visible simple closed curves no longer surround  $\mathcal{S}_i$
(and  $\mathcal{S}_1$ and $\mathcal{S}_2$); instead, it shrinks in average diameter.
At later times the third simple closed curve
is not visible in the frame; it is located further out compared to the scale used.  The bottom left and right panels of figure 1 indicate the behaviour of the level--$0$ sets of $\mathcal{D}_r$ at times $t = 8.0$ and $t = 9.0$,
which show a zoomed out view of the outer horizon.
For example, $t = 9.0$ shows part of the spherical GH that encloses both BHs.
These plots show that the level--$0$ sets of $\mathcal{D}_r$ could define a unique GH. However,
further investigation is needed, especially at later times when the outer MOTS
(the AH) is close to isolated again and has reached Kerr. This might require
going up to at least $t=20$ (where the horizon mass becomes constant), at which
time a GH might be present near the (outer) AH. At later times, however, when
the geometry is nearly Kerr and the spacetime is type {\bf{D}} everywhere, we may
also need to study the covariant derivative of the Weyl tensor.

\newpage

\section{Two merging equal mass non-spinning BHs}
Next we shall study a quasi-circular orbit of two merging, equal mass and non-spinning BHs.  In this simulation, which has not been presented elsewhere and the results of this simulation are consequently new, the Einstein toolkit infrastructure was used 
utilizing Brill-Lindquist initial data with BH positions and momenta set up to satisfy the ``QC-0'' initial condition \cite{QC0}.  
Instead of analyzing a sequence of MOTS throughout the merger, we seek to define and study a GH via the constant contours of ${\mathcal{D}}$ as they evolve through the merger.  
In the actual simulations, the real and imaginary parts of $I$ and $J$ are calculated using the Cartan (Weyl spinor) invariants, $\{{\bf \Psi}_i\}_{i=0}^5$, and the calculations are carried out using the orthonormal fiducial tetrad, as given by \cite{Lazarus}.  For comparison, in figures 2-4 we also plot the centroid and average radius of the MOTSs of the two initial BHs.  We have also verified that it is valid to approximate the MOTS as 1D spheres. We also note that in figures 2-4 the light blue curves are made of points corresponding to the exact MOTS, whereas the continuous pink curves are points on the spherically approximated MOTS.

\subsection*{Overview of Figures and Discussion} 
We study the algebraic properties of the Weyl tensor 
through a quasi-circular merger of two non-spinning, equal mass BHs.
The GH is theoretically identified by the level--$0$ set of the complex invariant ${\mathcal{D}}$.  
We find strong evidence that 
the level--$\varepsilon$ sets of $|{\mathcal{D}}|$ track a unique smooth GH through all stages of the BBH merger, 
by analyzing the time evolution of various level--$\varepsilon$ sets of $|{\mathcal{D}}|$  where, in particular, $\varepsilon_i  \equiv 3\times 10^{-4},\;5\times 10^{-4},\;1\times 10^{-3}$ for $(i=1-3)$, respectively. 
We note that
many additional figures (including those describing the inner and outer MOTS) are
presented in
\cite{Thesispaper}.

The figures  provide plots of various level sets of $|{\mathcal{D}}|$ 
as functions of $(x,y)\in \mathbb{R}^2$ at a fixed spatial coordinate value of $z = 0.3125$ and at selected instances of the time parameter, $t$, where $t = 0$ indicates the start of the numerical computation.  In each of the figures, the data corresponding to $x<0$ was obtained by rotating the data corresponding to $x>0$ by $180$ degrees about the $x = y = 0$ axis.  
In \cite{Thesispaper}, the contour plots of $|{\mathcal{D}}|$ (and ${\mathcal{D}}_r$ and other complex quantities) and 
its level--$0$ sets  and various other level--$\varepsilon$ sets
at all times $t = 0 - 42$ and in more detail and with magnified resolution were presented.  Here we shall illustrate the qualitative features of the contour plots of the  level--$\varepsilon$ 
sets $|{\mathcal{D}}|$ on a log scale  for $t = 12,\;16,\;20$.  
The overlaid green, red and white contours in each 
first panel (top left-hand side)
of the figures are the level--$\varepsilon_1$, level--$\varepsilon_2$ and level--$\varepsilon_3$ sets of $|{\mathcal{D}}|$, respectively.  The blue dots in the figures  give the centroids of the MOTSs of the two initial BHs as they evolve and serve to track the positions of these BHs through the merging process.  
At early times, when the configuration is close to 2 Kerr BHs and the spacetime is 
approximately of type {\bf{D}}, the time plots are somewhat unreliable.

At early times (e.g., at $t = 8$), each of the level--$\varepsilon$ sets are partitioned into pairs of simple disjoint closed curves, each of which contains the centroid of the MOTS of each of the two initial BHs.  At intermediate times (e.g., $t = 16$), the red level--$\varepsilon_2$ set and the white level--$\varepsilon_3$ set of $|{\mathcal{D}}|$ each form a third simple closed curve between the centroids of the MOTSs of the two initial BHs, which is centred at the origin (this occurs a little earlier for the green level-$\varepsilon_1$ set \cite{Thesispaper}).  At later times (e.g., at $t = 20$), for each respective $\varepsilon_i$, the multiple simple closed curves partitioning the level--$\varepsilon$ set of $|{\mathcal{D}}|$ have joined so that each level--$\varepsilon$ curve is now a single simple closed curve surrounding the 2 BHs.  It follows that the level--$\varepsilon_i$ curves at each $t$ form an invariantly defined, foliation invariant horizon that contains each separate BH at early times, and contains the merged BH at late times.     
In addition, we observe that the level--$\varepsilon_i$ contours are very close to each other, showing that the level--$\varepsilon$ sets vary continuously with $\varepsilon$.  We also observe that if $\varepsilon_I \leq \varepsilon_J$, then the 2D area enclosed by the level--$\varepsilon_I$ curve encloses the 2D area enclosed by the level--$\varepsilon_J$ curve, indicating that $|{\mathcal{D}}|$ decreases on average with average distance from the centroids of the MOTSs.

We compare the white level--$1\times 10^{-3}$ contours of $|{\mathcal{D}}|$ with points on the surface of the MOTSs of the 2 initial BHs at times $t = 12,\;16,\;20$
in the second panel (top right-hand side) of figures 2, 3 and 4.   The blue curves in the figures indicate the $(x,y)$ coordinates of a $z = 0.03125$ slice of the points on the ``spherically averaged" MOTSs of each initial BH (i.e.,  the 2D sphere centred at the centroid of the MOTS and whose radius is the average radius of the MOTS).  The centroid of this MOTS is again given as the central blue dot.  The light sky blue curves mark the $(x,y)$ coordinates along each ``actual" MOTS of each initial BH (i.e., without the spherical approximation).  
We note that the MOTSs of the two initial BHs are nearly spherically symmetric surfaces and that the white level--$1\times 10^{-3}$ set of $|{\mathcal{D}}|$ coincides closely with the MOTSs (both spherically averaged and exact), especially at early times.
This lends support to the choice of the white level--$1\times 10^{-3}$ sets of $|{\mathcal{D}}|$ as a representative approximation to the level--$0$ set of $|{\mathcal{D}}|$.  
In figure 4 for $t=20$ there is also an outermost MOTS (the AH) which is displayed in the final panel on a different scale.

The evolution of the level--$\varepsilon$ curves through the quasi-circular BBH merger in the figures is qualitatively similar to the sequence of MOTS that take place during the head-on collision simulation in \cite{ERIK} (see earlier) and the GH in the KT analysis \cite{KT}.  
However, our numerical  simulations 
were not run  to sufficiently late times 
to study the details of the bifurcation.

The complex invariant ${\mathcal{D}}$ vanishes if and only if its magnitude, $|{\mathcal{D}}| = 0$.  However, numerically finding the level--$0$ sets of $|{\mathcal{D}}|$ precisely is extremely difficult because $|{\mathcal{D}}|$ is a positive definite quantity, so that any numerical errors would provide a positive contribution.  Furthermore, in the numerical simulations it is possible that the actual zeros of $|{\mathcal{D}}|$ do not occur at any points which are sampled for the discrete mesh being used.  Thus, the level--$\varepsilon$ sets of $|{\mathcal{D}}|$ could indeed approximate the level--$0$ sets ${\mathcal{D}}$ for an appropriate preferred  $\varepsilon$-value.

If the value of $|{\mathcal{D}}|$ itself is small, then the locations of the local minima  of $|{\mathcal{D}}|$ could possibly indicate the positions of the actual zeros, which would be missed due to the numerical issues described above.  
It is conceivable that the GH conjecture should be modified so that the GH is defined as the set of points where $|{\mathcal{D}}|$ reaches the local minimum instead of being identically zero.  If this were the case, then locating the local minima of $|{\mathcal{D}}|$ would locate the GH precisely instead of approximating it.
However, further evidence from the anlysis of ${\mathcal{D}_r}$, as described later, perhaps suggests that this is not the case.

Thus, to further study the zeros of $|{\mathcal{D}}|$, which would indicate the zeros of ${\mathcal{D}}$, we study the positions of the local minima of $|{\mathcal{D}}|$.  To find these positions, it is sufficient to track the positions of the local minima of $|{\mathcal{D}}|$ along so-called ``slice plots" of $|{\mathcal{D}}|$ vs $y$ for a fixed $x$.  
[Along each slice plot, we find the values of $y = y_{min}$, where $|{\mathcal{D}}|$ assumes a local minimum value, and show that the point $(x_{min},y_{min})$ accurately represents the local minimum, and the  corresponding  points are then recorded.
We restrict our attention to finding the positions of the local minima of $|{\mathcal{D}}|$ whose corresponding $|{\mathcal{D}}|$ values lie within the range $\left[1\times 10^{-4},\;1.2\times 10^{-3}\right]$ which contains the set of values of $\varepsilon_i$ being considered for the level--$\varepsilon$ sets of $|{\mathcal{D}}|$.  ] 
Having done this for all possible fixed $x$, we then present plots of the locations of  these local minima of $|{\mathcal{D}}|$ with green dots  at selected times, $t =16$ and $20$, in panels 3 of figures 3 and 4.  These plots demonstrate 
that at time $t = 16$ (and earlier), 
the positions of the local minima of $|{\mathcal{D}}|$
appear to track the green level--$\varepsilon_1$ sets, while at time $t = 20$
(and later), these local minima appear to track more closely the white level--$\varepsilon_3$ sets.   Therefore, the figures show that the positions of the local minima of $|{\mathcal{D}}|$ accurately track the zeros of $|{\mathcal{D}}|$, and hence the GH,  during  the BBH merger. We will now argue that all local minima of $|{\mathcal{D}}|$ track closely the level--$0$ sets of ${\mathcal{D}}_r$.

The problem of estimating the level--$0$ sets of the complex ${\mathcal{D}}$ cannot be definitively resolved by analyzing $|{\mathcal{D}}|$ because, as previously noted, $|{\mathcal{D}}|$ is a positive definite quantity and the discrete resolution imposed by the numerical simulation does not necessarily allow one to accurately find level--$0$ sets of $|{\mathcal{D}}|$.  Thus, to gain further insight into the GH through the BBH merger, it is therefore helpful to analyze quantities which change sign through a zero.  This will qualitatively demonstrate the existence of level--$0$ sets.  
This avoids some of the problems of numerical resolution that occur when using the local minima of $|{\mathcal{D}}|$ to estimate the zeros of $|{\mathcal{D}}|$ (and hence ${\mathcal{D}}$).

In panels 4 of figures 3 and 4  we compare the level--$-0.01$ sets (in yellow) and the level--$+0.01$ sets (in lime green) of ${\mathcal{D}}_r$ at times $t = 16,\;20$
with the white level--$1\times 10^{-3}$ sets of $|{\mathcal{D}}|$.
[In these figures, the grey regions in each of the frames correspond to regions where $-0.01<{\mathcal{D}}_r<0.01$, as indicated in the colourbar, but the black regions correspond to regions where ${\mathcal{D}}_r\geq 1$  and the white regions correspond to regions where ${\mathcal{D}}_r\leq -1$].  
We know that a positive level set and a nearby negative level set indicates a sign change of the quantity being plotted.  Hence, there must be a surface bounded by the level $\pm0.01$ sets of ${\mathcal{D}}_r$  across which ${\mathcal{D}}_r$  changes sign.  This surface gives the level--$0$ set of ${\mathcal{D}}_r$.  By estimating the zeros of ${\mathcal{D}}_r$ in this manner, we reduce the possibility of numerical noise that comes with plotting the level--$0$ sets of ${\mathcal{D}}_r$ directly. 
Upon inspection of each figure, we see that the level--$\pm0.01$ sets of ${\mathcal{D}}_r$ are densely clustered around the centroids of the MOTSs of the initial BHs  and occur in close proximity with but are contained in the interior of the level--$1\times 10^{-3}$ sets of $|{\mathcal{D}}|$. 
It follows that there is strong evidence that the level--$1\times 10^{-3}$ set of $|{\mathcal{D}}|$ well approximates the elusive level--$0$ sets of the complex invariant ${\mathcal{D}}$.

We utilize  the notation described earlier (in section 2).
At around $t = 18.5 - 18.75$
the third MOTS $\mathcal{S}_i$  first appears and bifurcates to give rise to the fourth MOTS $\mathcal{S}_c$, and after 
$t=19$ these appear in the figures.
The outermost MOTS $\mathcal{S}_c$ (the AH) can be seen in the numerics but
is outside of the domain in the current figures. Indeed,
the outermost MOTS at late times is so big that the entire two dots and $S_1$ and $S_2$ and the inner $S_i$ are all squashed in the origin.
In figure 4 (panel 2)  for $t=20$,  the ``inner'' MOTS $\mathcal{S}_i$ is displayed as purple dots. 
The actual MOTS $\mathcal{S}_i$  as plotted is calculated exactly, and
is traced very well by the spherically averaged MOTS
(shown by the outer light pink oval figure, which overlays the purple dots
but is not clearly visible).
The outer MOTS  $\mathcal{S}_c$ is present, but is out of view and cannot be seen in the figure displayed.

\begin{figure}
    \centering
    \includegraphics[width = 12cm]{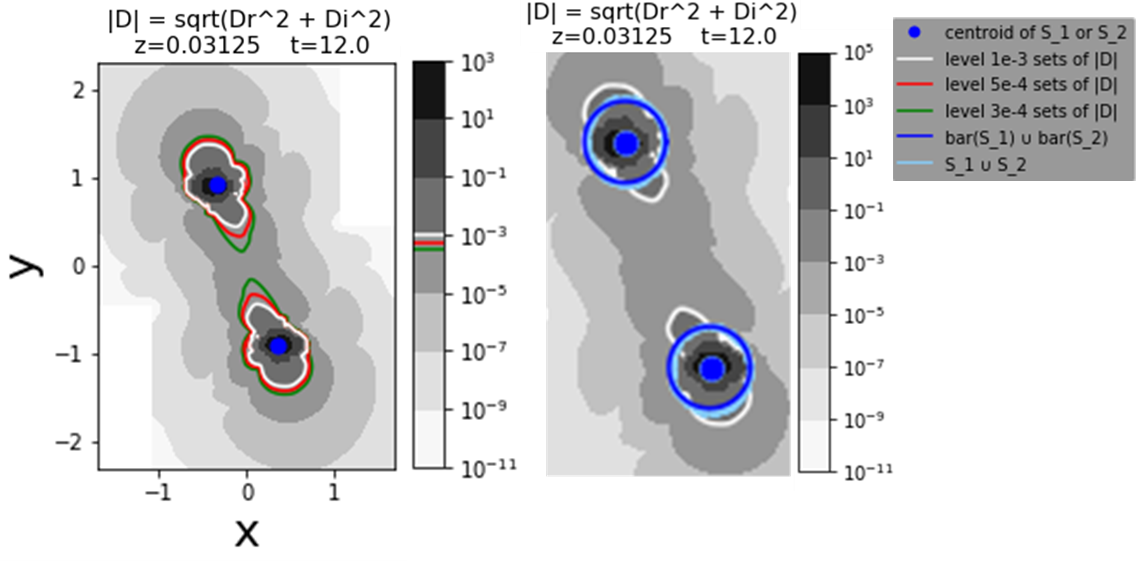}
    \caption{Plots of $|D| = \sqrt{x^2 + y^2}$ at time $t = 12$ from the quasi-circular orbit of two merging, equal mass and non spinning BHs as functions of $x$ and $y$ for fixed $z = 0.03125$.  The right panel 2 has the same horizontal and vertical scale as the left panel 1 but the plot is taken at magnified resolution.}
    \label{QCO12}
\end{figure}

\begin{figure}
    \centering
    \includegraphics[width = 12cm]{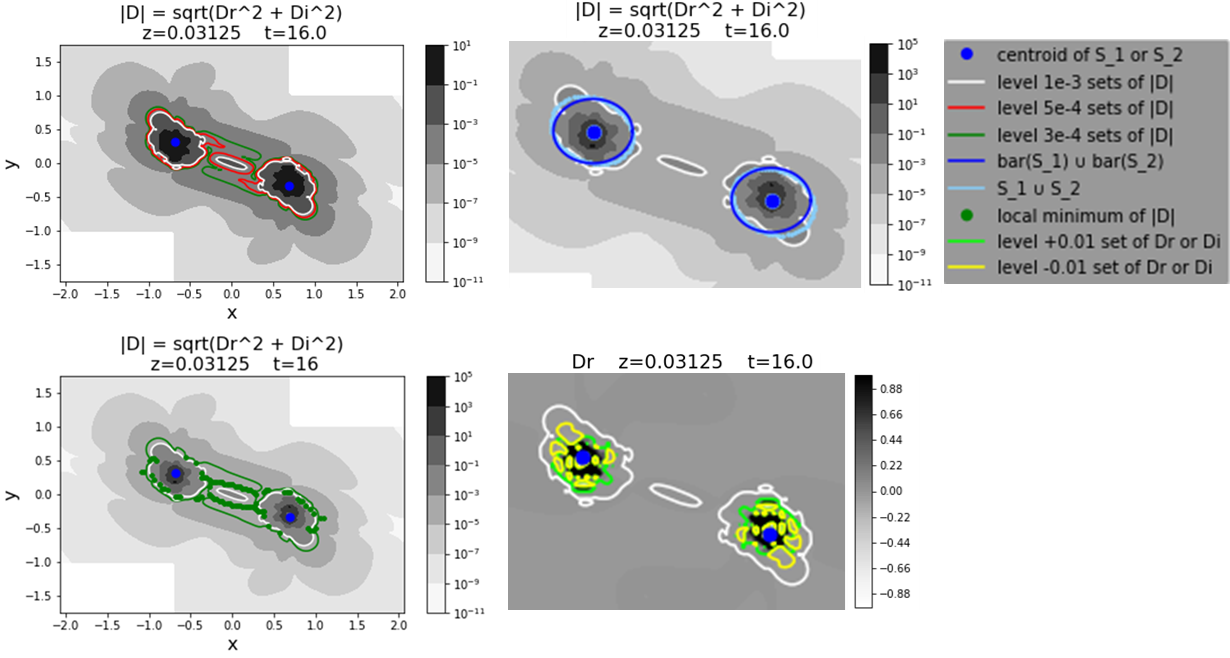}
    \caption{Upper left, lower left (panel 3) and upper right panels: plots of $|{\mathcal{D}}| = \sqrt{x^2 + y^2}$ at time $t = 16$ from quasi-circular orbit of two merging, equal mass and non spinning BHs as functions of $x$ and $y$ for fixed $z = 0.03125$.  Lower right panel 4: Plot of ${\mathcal{D}}_r$ at time $t = 16$ from the quasi-circular BBH merger.  In this figure and in Figure 4, all 4 plots are taken to be at the same scale but the lower and upper right plots are at increased resolution.}
    \label{QCO16}
\end{figure}

\begin{figure}
    \centering
    \includegraphics[width = 12cm]{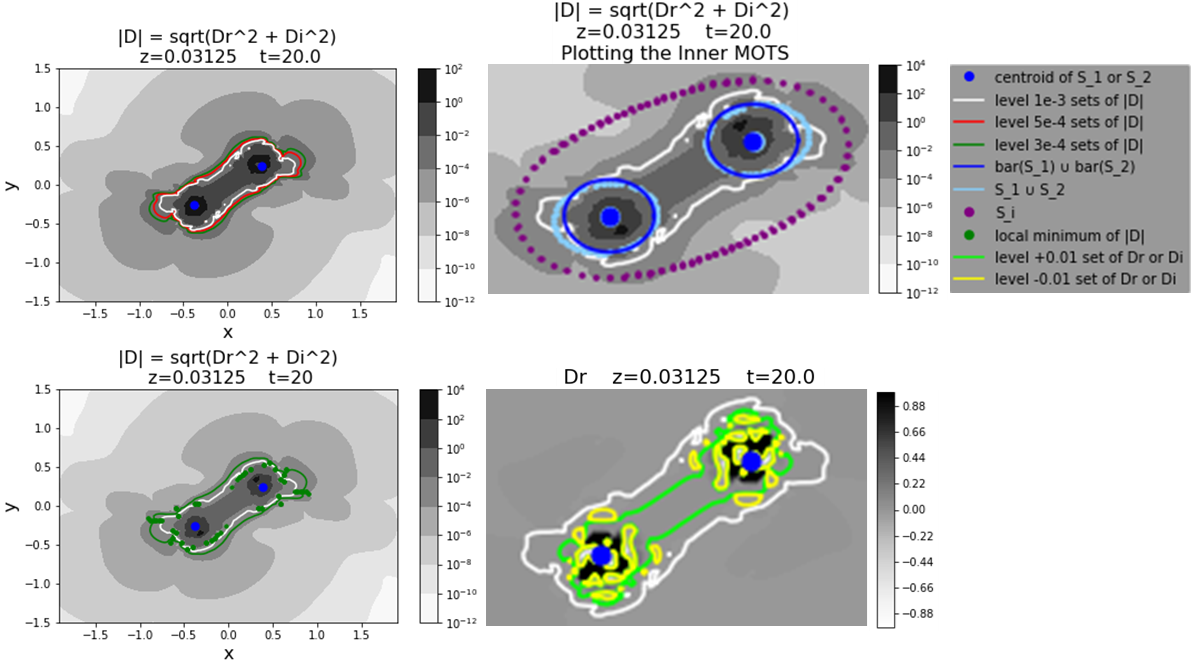}
    \caption{Upper left, lower left and upper right panels: plots of $|{\mathcal{D}}| = \sqrt{x^2 + y^2}$ at time $t = 20$ from the quasi-circular orbit of two merging, equal mass and non-spinning BHs as functions of $x$ and $y$ for fixed $z = 0.03125$.  Lower right panel 4: Plot of ${\mathcal{D}}_r$ at time $t = 16$.}
    \label{QCO20}
\end{figure}

\newpage

\section{Conclusions} 

We have studied the conjecture that \cite{CMS}
if a BH spacetime is zeroth-order algebraically general, a 
geometric horizon (GH) in which the spacetime is algebraically special,
and is identified by the vanishing of the complex scalar invariant ${\mathcal{D}}$,
characterizes a smooth foliation invariant surface (horizon) associated with the BH.

We have evaluated this conjecture by studying the GH  numerically
in the two physically relevant  situations of the head-on collision of unequal mass BHs
and two merging, equal mass and non-spinning BHs.
In the head-on collision simulation
${\mathcal{D}}_i= 0$, and so we studied the level--$0$ sets of ${\mathcal{D}}_r$,
which reveals the existence of a MOTS which plausibly defines the GH.

We then studied the algebraic properties of the Weyl tensor 
through a quasi-circular merger of two non-spinning, equal mass BHs.
The GH is theoretically identified by the level--$0$ set of the complex invariant ${\mathcal{D}}$, which we study via the level--$\varepsilon$ sets of $|{\mathcal{D}}|$, and we find strong numerical evidence that 
these level--$\varepsilon$ sets identify a unique smooth GH through all stages of the  binary black hole merger (as displayed in the figures). Due to the problems associated with finding the zeros of a positive-definite quantity with numerical errors,
we support our findings by also studying the local minima of $|{\mathcal{D}}|$ and the level--$0$ sets of $\text{Re}({\mathcal{D}})$.

There are a number of things to do in future work. In the 
head on collision simulations 
it is of interest to study late times in more detail when the the AH
has reached Kerr and is again close to being isolated.
In the BBH merger more resolution is desirable, and the problem 
of two unequal BH masses is of interest. In both cases it would also be of interest to
study the behaviour of the covariant derivative of the Weyl tensor.
This might be facilitated by computing MOTS in an
(algebraically) "preferred" null frame.

\newpage

\section*{Acknowledgements}  

This work was supported financially by NSERC (AAC and ES). We would like
to thank Ivan Booth and Daniel Pook-Kolb for discussions. JMP would also like
to thank the Perimeter Institute for Theoretical Physics for hospitality during
this work. Research at Perimeter Institute is supported, in part, by the Government of Canada through the Department of Innovation, Science and Industry
Canada and by the Province of Ontario through the Ministry of Colleges and
Universities.


\bibliographystyle{plain}
\bibliography{sources.bib}

\end{document}